\documentclass[%
reprint,
amsmath,
amssymb,
amsfonts,
aps,
pre,
floatfix,
]{revtex4-2}

\usepackage{graphicx}
\usepackage{dcolumn}
\usepackage{bm}
\usepackage{nicefrac}
\usepackage[capitalise]{cleveref}
\crefname{app}{appendix}{appendices}
\usepackage[mathlines]{lineno}
\usepackage{xcolor}

\renewcommand{\d}{\;\mathrm{d}}

\newcommand{\deriv}[3][]{\frac{\mathrm{d}^{#1}#2}{\mathrm{d}#3^{#1}}}

\DeclareMathOperator{\intinf}{\int_{-\infty}^{\infty}}
\DeclareMathOperator{\re}{Re}
\DeclareMathOperator{\im}{Im}

\DeclareMathOperator{\fluc}{\psi}
\DeclareMathOperator{\lorentzsig}{\gamma}

\newcommand{\density}{\rho}

\newcommand{\gav}[1]{\left\langle #1 \right\rangle}

\newtheorem{theorem}{Theorem}

\begin{document}

\title{Binary synchronization of noise-coupled oscillators} 

\author{Jeremy Worsfold}
\affiliation{Centre for Mathematical Biology, Department of Mathematical Sciences, University of Bath, Claverton Down, Bath BA2 7AY, United Kingdom}
\author{Tim Rogers}
\affiliation{Centre for Mathematical Biology, Department of Mathematical Sciences, University of Bath, Claverton Down, Bath BA2 7AY, United Kingdom}

\begin{abstract}
After decades of study, there are only two known mechanisms to induce global synchronization in a population of oscillators: deterministic coupling and common forcing. The inclusion of independent random forcing in these models typically serves to drive disorder, increasing the stability of the incoherent state. Here we show that the reverse is also possible. We propose and analyse a simple model of purely noise coupled oscillators whose linear response around incoherence is identical to that of the paradigmatic Kuramoto model, but which exhibits binary phase locking instead of full coherence. We characterise the phase diagram, stationary states and approximate low dimensional dynamics for the model, revealing the curious behaviour of this new mechanism of synchronization. 
\end{abstract}


\maketitle


\section{Introduction}

Emergent synchronization has been studied extensively over the last half century, initiated by Kuramoto's introduction in 1975 \cite{Kuramoto1975,kuramotoreview} of a paradigmatic model of globally coupled oscillators. 
Numerous applications exist, from power networks \cite{powernetworks,powernetworks2}, to Josephson arrays \cite{wanatabe-strogatz}, synchronization of fireflies \cite{ermentrout1991adaptive} and bacterial suspensions \cite{WeakSynch2017}. Most modern versions of Kuramoto's model feature two sources of randomness: the quenched disorder of the randomly chosen intrinsic frequencies, and independent constant-coefficient stochastic noise terms in the dynamics of the oscillators. The first of these models natural variability in populations, the second models inherent stochasticity or unpredictability in the behavior of individual elements. Invariably, both are drivers of global disorder acting counter to the deterministic coupling, raising the coupling strength required to induce synchronization and lowering the coherence of the emergent states.

In other areas of physics situations have been observed in which randomness is in fact a driver global ordering. For example, equilibrium statistical physics possesses many examples of entropically driven ordered states which can be thought of as emerging from purely random interactions, a canonical example being Onsager's work on nematic fluids \cite{Onsager1949}. Recently, there has been some effort to search for similar effects in the dynamics of coupled oscillators. Promising work has included studies considering common noise terms --- for example arising from environmental fluctuations --- that aid synchronization \cite{commonnoise,sinenoisemodel}, but so far the possibility of independent noise driving the emergence of coherent states has been overlooked. 

\begin{figure}
    \centering
    \includegraphics[width=0.48\textwidth]{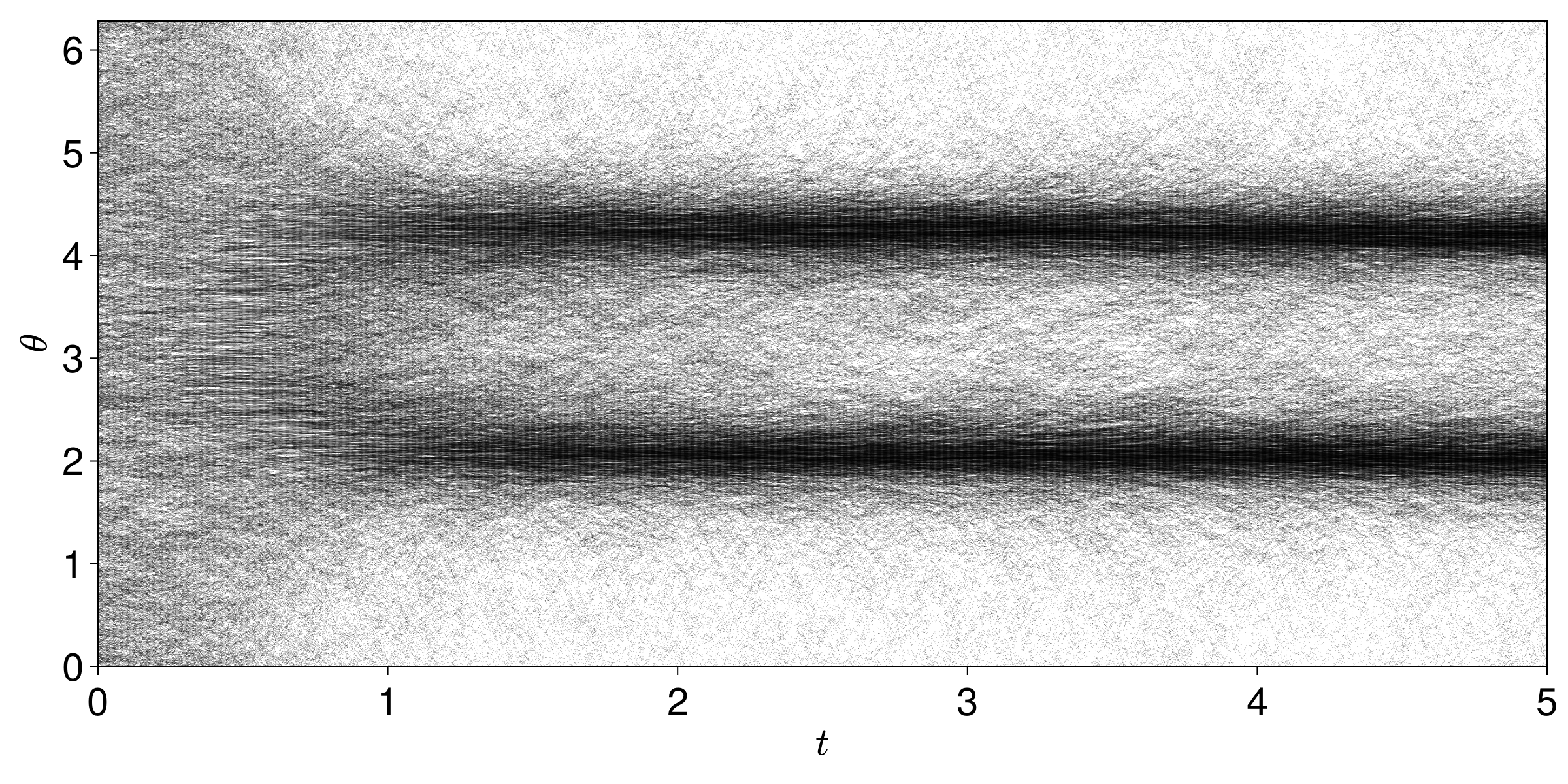}
    \caption{Emergence of binary synchronization from a sample of $N=2\times10^3$ oscillators for the Cauchy noise case ($\alpha=1$) of our model and Lorentz distributed frequencies with $\kappa=5,\;\gamma=0.1$.}
    \label{fig:particletrace}
\end{figure}

Here, we show that in fact the phase diagram of the Kuramoto model can be replicated in a population of oscillators with purely random forcing. As in the original Kuramoto model the oscillators are only influenced by their phase difference to the others oscillators.
Since only the strength of the noise changes, there is no bias on the direction the oscillator moves; remarkably, we show this can be sufficient to induce similar features to traditional Kuramoto coupling models. We further show that the emergent behaviour such as the steady states and individual oscillator movement can be characterised in the order regime, which exhibits a curious phenomenon of binary synchronisation, see \cref{fig:particletrace}. We derive explicit expressions for the steady states and capture the qualitative behaviour of the order parameters with approximate low dimensional dynamics. 

\section{Model}

We consider a population of $N$ oscillators with phases $\theta_n(t)$.
Each oscillator has an inherent natural frequency, $\omega_n$, sampled from a distribution $g(\omega)$, which should be considered as a source of quenched disorder.
There is no deterministic coupling, but each oscillator will be subject to an independent Levy noise term, $\xi_n(t)$, whose strength at time $t$ is determined by summing contributions from the rest of the population. Specifically, we write 
\begin{align}
    \dot{\theta}_n = \omega_n + \bigg(\frac{1}{N}\sum_{m}f(\theta_n-\theta_m)\bigg)^\beta  \xi_n(t)\,,
    \label{wors}
\end{align}
where $f$ is a function to be chosen, and $\xi_n$ is a Levy noise with index $\alpha=2/\beta$. Two cases are of particular interest: if $\beta=1$ then we have Gaussian white noise; if $\beta=2$ then noise terms are Cauchy distributed which facilitates the computation of low-dimensional dynamics~\cite{lowDimCauchy}.

Our main object of study will be the oscillator density $\rho(\theta,\omega,t)=\frac{1}{Ng(\omega)}\sum_n\delta(\theta-\theta_n(t))\delta(\omega-\omega_n)$. Using the shorthand $\gav{\cdots}=\int (\cdots)g(\omega)\d\omega$ to denote averaging over the distribution in intrinsic frequencies, the noise strength term in (\ref{wors}) {can be written simply as the convolution $\gav{\rho \ast f}$.
Applying standard methods~\cite{worsfold2022density}, one can then take the limit $N\to\infty$ to obtain an integro-differential equation for the oscillator density:
\begin{align}
    \partial_t\rho = - \omega\partial_\theta \rho+\partial^{\alpha}_{|\theta|}\left(\rho\,\gav{\big.\rho \ast f }^2 \right).
    \label{eq:generaldensityevolution}
\end{align}

Here we have used the Riesz derivative $\partial^\alpha_{|\theta|}$, defined through its action under Fourier transformation Specifically, $\int_{-\pi}^\pi e^{-ik\theta}\partial_{|\theta|}^\alpha u(\theta)\d \theta=-|k|^\alpha u_k$. Note that in the Gaussian ($\alpha=2$) case, this is simply the diffusion operator. By symmetry, the above equation admits a fixed-point solution that does not vary in phase or time, $\rho_\circ\equiv 1/2\pi$, known as the incoherent state. This state may or may not be stable. The phenomenon of synchronisation may broadly be defined as the emergence of one or more peaks in the oscillator phase density which persist over time. An indicator of synchronisation in the system is provided by the complex order parameter 
\begin{align}
    z= \int_{-\pi}^\pi\gav{\rho e^{i\theta}}\d \theta\,.
    \label{eq:orderparameter}
\end{align}
The argument of $z$ gives the average phase, while the modulus describes the level of global coherence; in the incoherent state we have $|z|=0$, whilst full synchronisation implies $|z|=1$.

We begin by examining the dynamics of fluctuations around the incoherent state, studied in detail in \cite{KuramotoFiniteScaling}. In doing so, we will identify a choice of noise-coupling function $f$ that exactly maps the fluctuations in our system to those of the well-studied noisy Kuramoto model. If $\rho=\rho_\circ + \varepsilon\fluc$, where $\varepsilon$ is small, then to leading order \eqref{eq:generaldensityevolution} yields
\begin{align*}
    \partial_t\fluc = - \omega\partial_\theta\psi +f_0\partial^\alpha_{|\theta|}\fluc+ 2\rho_\circ\gav{\psi\ast \partial^{\alpha}_{|\theta|}f}  \,,
\end{align*}
Performing the same analysis on the noisy Kuramoto model yields a similar equation for the linear evolution of fluctuations. In fact, if we make the choice $f(\theta) = 1- \kappa \cos(\theta)$ then we obtain precisely the same expression for both models:
\begin{align}
    \partial_t\psi & = - \omega\partial_\theta\psi +\partial^{\alpha}_{|\theta|}\psi +2\kappa \rho_\circ\gav{\psi\ast\cos} \,.
    \label{eq:kuramotostability}
\end{align}
In Appendix \ref{app:stabincoherent} we give full details of the derivation of this result for both models, and show that this choice for the noise coupling function is the only one for which the statistics fluctuations match. 

\section{Linear stability at incoherence}

As a consequence of the equivalence of our noise-coupled oscillator model with the noisy Kuramoto model, the systems have identical phase boundaries for the onset of synchronisation. 
Following Strogatz and Mirollo \cite{strogatz1991}, we show that for specific choices of the frequency distribution the exact stability boundary for the homogeneous state can be calculated. In \cite{strogatz1991}, it is shown that only the first Fourier mode of the perturbation, $\fluc_1(\omega)$ need be considered in the stability analysis. Briefly, this can be seen as the higher modes do not have any contributions from the final term while the first term is diffusive and thus all higher modes decay over time. For the first Fourier mode, we have
\begin{align*}
    \partial_t \fluc_1 = (i\omega-1)\fluc_1 + \kappa\gav{\fluc_1}.
\end{align*}
Assuming this Fourier mode has an exponential form: $\fluc_1(\omega)=\phi(\omega)e^{\eta t}$, then
\begin{align*}
    \eta\phi(\omega) = (i\omega-1)\phi(\omega)+\kappa\gav{\phi}.
\end{align*}
The average over frequencies, $\gav{\phi}$ is just a constant and so the frequency dependence of $\fluc_1$ is 
\begin{align*}
    \phi(\omega) = \frac{\kappa\gav{\phi}}{\eta+1-i\omega}.
\end{align*}
In addition, the average must be self-consistent so that 
\begin{align*}
    \gav{\phi} = \intinf\frac{\kappa\gav{\phi}}{\eta+1-i\omega}\d\omega
\end{align*}
or equivalently $1 = \kappa\gav{1/(\eta+1-i\omega)}.$ It can be shown that if $g(\omega)$ is a non-increasing function for $\omega>0$, and is symmetric about the origin, then there exists at most one solution for $\eta$ and it is necessarily real (see \cite{strogatz1991} and \cite{Mirollo1990}). Hence, we need only take the real component 
\begin{align}
    1 = \kappa\gav{\frac{1+\eta}{(1+\eta)^2+\omega^2}}
    \label{eq:disorderavstab}
\end{align}
as the symmetry in $\omega$ implies that the imaginary component integrates to zero.

Here we only show the result for Lorentz distributed frequencies $g(\omega)=(\gamma^2/\pi)[\gamma^4+\omega^2]^{-1}$, with width $\gamma^2$ as it is the focus of subsequent sections. The integrand in \eqref{eq:disorderavstab} can now be separated into partial fractions and integrated with standard results to give 
\begin{align*}
    \intinf\frac{1+\eta}{(1+\eta)^2+\omega^2}g(\omega) \d\omega  = \frac{1}{1+\gamma^2+\eta}\;.
\end{align*}
Comparing this to \eqref{eq:disorderavstab}, it is clear that $\kappa = 1+\gamma^2+\eta$. The system is stable if $\eta<0$, which we deduce is satisfied when
\begin{align}
    \kappa < 1 + \gamma^2
\end{align}
which can be seen to match with simulations for various values of $\kappa$ and $\gamma$ in \cref{fig:lorentzstationary}.

\section{Stationary State without disorder}

With Lorentz distributed intrinsic frequencies, $g(\omega;\gamma) = (\lorentzsig^2/\pi)\left[\omega^2+\lorentzsig^4\right]^{-1}$, the incoherent state is stable for $\kappa<1+\lorentzsig^2$, as shown in \cref{fig:lorentzstationary} (a). 

Although the dynamics of our model are indistinguishable from the Kuramoto model in the incoherent phase, the behaviour on the other side of the phase transition is dramatically different. 
As illustrated in \cref{fig:particletrace} and \cref{fig:lorentzstationary} (b-c) simulations of our model exhibit binary synchronisation, with the oscillator population spontaneously dividing into two quasi-coherent phase-locked groups with a consistent separation distance between groups. 
The remainder of the paper will be devoted to studying this unusual behaviour.

The starting point for all our analysis will be the Fourier representation of the governing equation \eqref{eq:generaldensityevolution}. Writing $\rho_k$ for the $k^{\text{th}}$ Fourier mode of $\rho$ (note that $z=2\pi\gav{\rho_{-1}}$), we have
\begin{equation}
\begin{aligned}
    \partial_t\rho_k  =&- ik\omega \rho_k - |k|^{\alpha}\rho_k + |k|^{\alpha}\kappa\big(  z\rho_{k-1}+\bar{z}\rho_{k+1} \big) \\ 
    & - |k|^{\alpha}\frac{\kappa^2}{4}\left(\bar{z}^2\rho_{k+2}+2|z|^2\rho_k+z^2\rho_{k-2}\right)\,.
\label{eq:rhoevolution}
\end{aligned}
\end{equation}

First, we characterize the state of full binary synchronization occurring when $\kappa>1$ if the oscillators all have the same intrinsic frequency. We pick an appropriately chosen rotating reference frame such that the density is symmetric and centred at zero. Then \eqref{eq:rhoevolution} simplifies to 
\begin{equation}
\begin{aligned}
\rho_k = & -|k|^\alpha\bigg(\rho_k - |z|\kappa\big(  \rho_{k-1}+\rho_{k+1} \big) \\ & + \frac{|z|^2\kappa^2}{4}\left(\rho_{k+2}+2\rho_k+\rho_{k-2}\right)\bigg)\,,
\label{eq:rhostationarydelta}
\end{aligned}
\end{equation}
where we can further identify $|z|=2\pi\rho_1$. This infinite system of equations can be collapsed by making the ansatz $\rho_k=T_k(\cos(\Delta))/2\pi$, where $T_k$ is the $k^{\text{th}}$ order Chebyshev polynomial of the first kind and $\Delta$ is a non-negative number. Collapsing \eqref{eq:rhostationarydelta} is possible since the Chebyshev polynomials obey the following rules:
\begin{align*}
    T_{k+1}(x) + T_{k-1}(x) &= 2T_1(x)T_k(x), \\
    T_{k+2}(x) + T_{k-2}(x) &= 2\left(2T_1(x)^2-1\right)T_k(x).
\end{align*}
Substituting our ansatz into \eqref{eq:rhostationarydelta} and writing $T_k(\cos\Delta)=T_k$ for brevity, this becomes
\begin{align*}
    0 & = |k|^\alpha\left( T_k -2\kappa T_1^2T_k + \frac{\kappa^2}{4}T_1^2\left(4T_1^2T_k\right)\right) \\
    & = |k|^\alpha T_k\left(1-\kappa T_1^2\right)^2\;.
\end{align*}
Thus the ansatz is a solution if $T_1 = \kappa^{-1/2}$, from which we deduce $\Delta=\arccos(1/\sqrt{\kappa})$. 

\begin{figure}
    \centering
    \includegraphics[width=8.6cm]{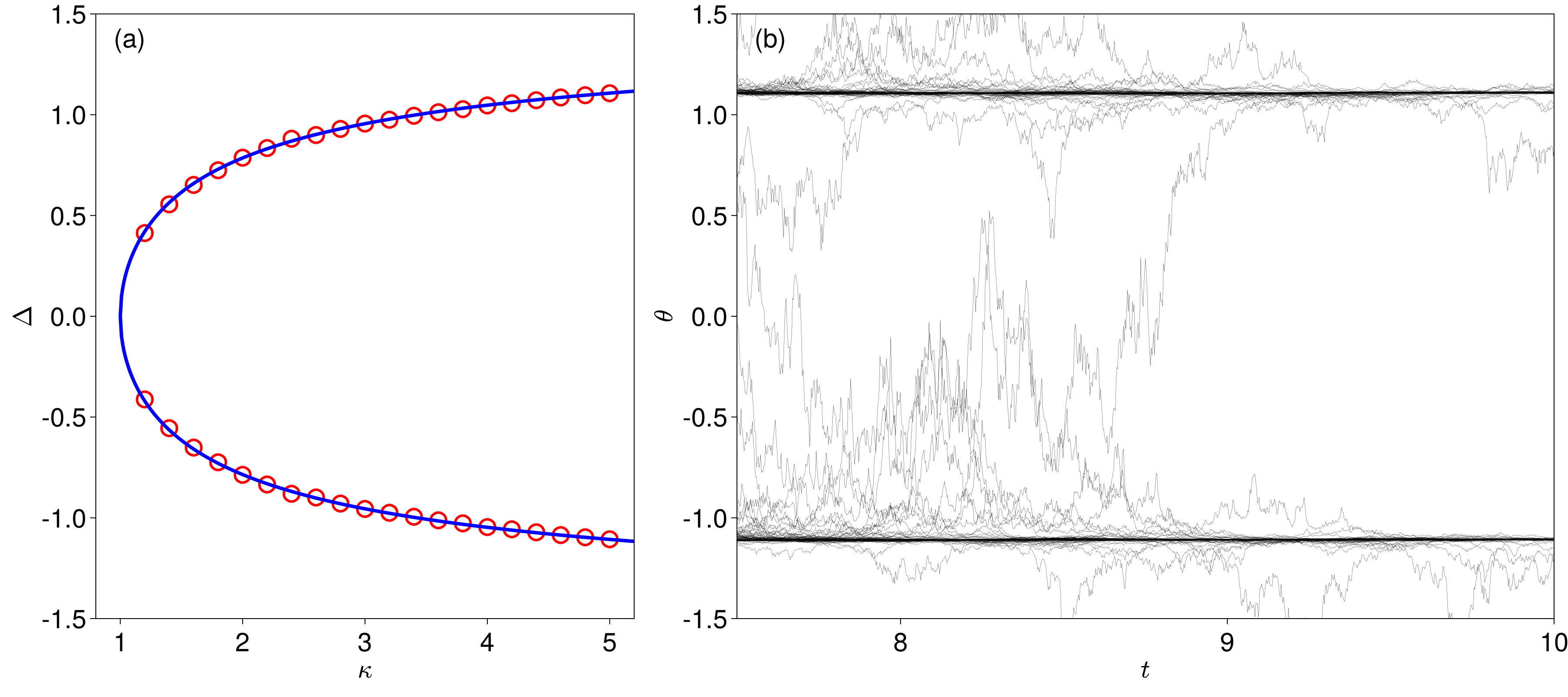}
    \caption{Exact binary synchronization without disordered intrinsic frequencies. (a) The half-separation of the peaks, $\Delta$, as a function of $\kappa$: the line shows theoretical steady state; open circles show simulation results. (b) Simulation close to the binary synchronised state with $\kappa=5$ for Brownian noise. Stray oscillators diffuse across the gap between the two peaks.}
    \label{fig:strayparticle}
\end{figure}

This solution corresponds to the oscillator phase density condensing to a symmetric pair of Dirac masses with separation $2\Delta$. That is, $\density(\theta) = \left(\delta(\theta-\Delta) + \delta(\theta+\Delta)\right)/2$. The oscillators become phase locked in these binary positions as the contributions to from each Dirac mass to the convolution term in \eqref{eq:generaldensityevolution} negatively interfere to precisely cancel each other. 

In \cref{fig:strayparticle}, while the system approaches the two-peaked steady state, erratic particles diffuse from near one peak to the other. 
To comprehend better this steady state for identical oscillators, we study the behaviour of a single stray oscillator in the Brownian noise case ($\alpha=2$). 
Consider the motion of this stray oscillator to be between the two peaks (i.e. $\theta\in[-\Delta,\Delta]$) governed by 
\begin{align}
    \dot \theta = \left[1 - \sqrt{\kappa} \cos(\theta)\right] \xi(t)\;. \label{eq:strayparticle}
\end{align}
The expected time, $\tau(\theta)$, to reach a distance $\epsilon$ from one of the peaks is the solution to $\tau''(\theta) = \left[1-\sqrt{\kappa}\cos(\theta)\right]^{-2}$ subject to the boundary conditions $\tau(\Delta-\epsilon)=\tau(\epsilon-\Delta)=0$. 
Interestingly, the equation above is the same if the oscillators starts at $\Delta<|\theta_0|<\pi$ since the increase in noise strength is matched by the larger distance from the peaks. 
We also observe that, since there is no drift term in \cref{eq:strayparticle}, the probability the stray oscillator will reach one peak as opposed to the other is directly proportional to its distance from the peak relative to the other. 
Explicitly, we have that $p_{\Delta}(\theta_0)=(\Delta+\theta_0)/2\Delta$ and $p_{-\Delta}(\theta_0)=1-p_{\Delta}(\theta_0)$ where $p_{\pm\Delta}(\theta_0)$ are the probabilities the oscillator will reach $\theta=\pm\Delta$ eventually (see \cite{gardiner} and Appendix \ref{app:stochstability}).

Studying an individual oscillator gives intuition for why the two-peaked state is stable for $\kappa>1$. Assume all particles are perturbed by a small amount, $\vartheta_i$. If the perturbation is small enough so that $\sqrt{\kappa'}=\kappa|z|>1$ still holds, each oscillator still has a solution to $\dot \theta = [1-\sqrt{\kappa'}\cos(\theta)]\xi(t)$ and the probability that it will return to its closest peak can be approximated by $p_\Delta(\Delta-\vartheta_i)$. Hence, at least close to this solution, it appears to be stochastically asymptotically stable \cite{MaoStochStability}. Due to the irregularity of the two-peak solution, formally showing stability from the macroscopic scale would be a more involved task, which we leave for future work. 

\section{Stationary state with disorder}

We broaden our investigation now to address the more general case of heterogeneous intrinsic frequencies. In the last two decades, great strides have been made in describing the dynamics of Kuramoto-like systems in terms of simple equations for the order parameters. Starting with the Watanabe-Strogatz variables \cite{wanatabe-strogatz}, it was shown that a suitable transformation on the oscillator phase to a homogeneous, stationary phase results in just three equations needed to describe the full dynamics of the system of $N$ particles \cite{strogatzMobius}. Ott and Antonsen \cite{ott2008low} subsequently derived similar equations for the order parameters. 

These equations connect the nonequilibrium transition from the incoherent state to the synchronised state.
Incorporating intrinsic noise has presented another challenge as the Ott-Antonsen manifold no longer holds when the oscillator phases have additive Brownian noise \cite{tyulkina2018dynamics}. 
When Cauchy noise is included instead, it has been shown to give equivalent low dimensional dynamics as systems with Lorentz distributed frequencies \cite{cauchynoise}. 
Exact low dimensional expressions for the steady states of models more complex coupling has also been achieved with Cauchy noise \cite{lowDimCauchy}. 
We use a similar approach here to identify the non-trivial steady state of the model presented above.

When the incoherent state is unstable, the distribution evolves towards a bimodal distribution (see \cref{fig:particletrace}) which is surprisingly distinct to the unimodal distribution seen in the Kuramoto model. As with the majority of Kuramoto-based models, we examine the case where the oscillator frequencies are Lorentz distributed. This enables the bimodal distribution at the steady state of \eqref{eq:rhoevolution} to be solved exactly. 

In steady state, the mode-coupling equation \eqref{eq:rhoevolution} is of a class studied by T\"onjes and Pikovsky~\cite{cauchynoise}, who proposed to seek solutions of the form
\begin{align}
    \rho_k(\omega) = c_1\lambda_1(\omega)^k +c_2\lambda_2(\omega)^k\;,
    \label{eq:generalansatz}
\end{align}
where $\lambda_1,\lambda_2$ are complex parameters lying in the unit disc, and $c_1,c_2$ are normalization coefficients summing to $1/2\pi$. Explicitly, 
\begin{align}
    c_1(\omega) & = \frac{1}{2\pi}\left( 1 - \frac{\lambda_2(1-|\lambda_1|^2)(1-\overline{\lambda_2}\lambda_1)}{\lambda_1(1-|\lambda_2|^2)(1-\overline{\lambda_1}\lambda_2)} \right)^{-1}
\end{align}
and $c_2(\omega)=1/2\pi - c_1(\omega)$.
In real (phase) space, this ansatz posits that the density of oscillators with a given frequency has the form of the product of two wrapped Cauchy distributions, also known as a Kato-Jones distribution (cf. \cite{KatoJones,cauchynoise}). Specifically, this distribution can be written as
\begin{align*}
    \rho(\theta,\omega) = \frac{1}{2\pi M} \prod_{n=1}^2 \frac{1-|\lambda_n|^2}{|e^{i\theta}-\lambda_n|^2}\;,
\end{align*}
where $M$ is a normalisation constant given by
\begin{align*}
    M = \frac{\lambda_1(1-|\lambda_2|^2)}{(\lambda_1-\lambda_2)(1-\bar{\lambda}_2\lambda_1)} + \frac{\lambda_2(1-|\lambda_1|^2)}{(\lambda_2-\lambda_1)(1-\bar{\lambda}_1\lambda_2)}
\end{align*}
and we have omitted the dependence of $\omega$ on $\lambda_1,\lambda_2$.
In this distribution, the arguments of the complex parameters $\lambda_1$ and $\lambda_2$ determine the positions of the two peaks, while the moduli determine the coherence and relative weighting of the peaks. 

Applying \eqref{eq:generalansatz} to \eqref{eq:rhoevolution} reveals an equation that must be satisfied by each of these complex parameters. Firstly, for $k\geqslant1$
\begin{widetext}
\begin{align*}
    0  = \sum_{n=1}^2 c_n\bigg\{-ik\omega\lambda_n^k - |k|\Big[\lambda_n^k - \kappa z\big(  \lambda_n^{k-1}+\lambda_n^{k+1} \big) + \frac{\kappa^2 z^2}{4}\left(\lambda_n^{k-2}+2\lambda_n^k+\lambda_n^{k+2}\right)\Big]\bigg\}\;.
\end{align*}
\end{widetext}
If the argument in the curly brackets is zero for each of $n=1,2$ then dividing through by $k\lambda_n^k$, we obtain
\begin{align}
    0 & = i\omega + \left[1 - \frac{\kappa z}{2}(\lambda+\lambda^{-1})\right]^2\;.
    \label{eq:omegazlambda}
\end{align}
Recalling that $z=2\pi\gav{\rho_{-1}}$, we see that this equation must be solved self-consistently with $z$ determined as a function of $\lambda_{1,2}$. Here we appeal to the remarkable result of Ott and Antonsen \cite{ott2008low}, that when the intrinsic frequencies are chosen from a Lorentz distribution, disorder averaging can be replaced by evaluation at a particular complex frequency. Specifically, if $g(\omega)=\left[(\omega-i\lorentzsig^2)^{-1} - (\omega+i\lorentzsig^2)^{-1}\right]/2\pi i$ then $\gav{\rho_k}=\rho_k(-i\gamma^2)$. The symmetry of $g$ implies that we may choose a frame of reference in which the disorder-averaged stationary distribution $\gav{\rho}$ is also symmetric, implying that $\gav{\lambda_1}=\gav{\overline{\lambda_2}}=\lambda$. The frequency-averaged distribution can thus be written as
\begin{align}
    \gav{\rho^{\text{st}}(\theta)} = \frac{1}{2\pi}\frac{1-|{\lambda}|^2}{1+|{\lambda}|^2}\left(\frac{|1-{\lambda}^2|}{|e^{i\theta}-{\lambda}||e^{i\theta}-\bar{\lambda}|}\right)^2\,.
    \label{eq:stationarydistr}
\end{align}
Moreover, the frequency averaged Fourier modes, $\gav{\rho_k} = \gav{c_1}\gav{\lambda}^k + \gav{c_2}\gav{\overline{\lambda}}^k$, now have normalisation constants given by
\begin{align}
    \gav{c_1} & = \frac{1}{2\pi}\frac{\lambda(1-\bar{\lambda}^2)}{\lambda(1-\bar{\lambda}^2)-\bar{\lambda}(1-\lambda^2)}
\end{align}
or, after some manipulation
\begin{align*}
    \gav{c_1} = \frac{1}{4\pi}\left(1+ i\frac{|\lambda|^2-1}{|\lambda|^2+1}\frac{\re(\lambda)}{\im(\lambda)}\right)\, ,
\end{align*}
from which we see that $\gav{c_1}=\gav{\overline{c_2}}$ recalling that $\gav{c_2}=1/2\pi-\gav{c_1}$.
This is also apparent as we require $\gav{\rho_k}=\gav{c_1}\lambda^k+\gav{c_2}\bar{\lambda}^k$ to be real which is only satisfied for all $k$ if $\gav{c_1}=\gav{\overline{c_2}}$.
Returning to the order parameter, $z = 2\pi(\gav{c_1}\lambda + \gav{c_2}\bar{\lambda})$, after some simplification we can compactly express this as 
\begin{align}
    z = \frac{\lambda + \bar{\lambda}}{|\lambda|^2+1}\,.
    \label{eq:zintermsoflambda}
\end{align}

\begin{figure}[t]
    \centering
    \includegraphics[width=8.6cm]{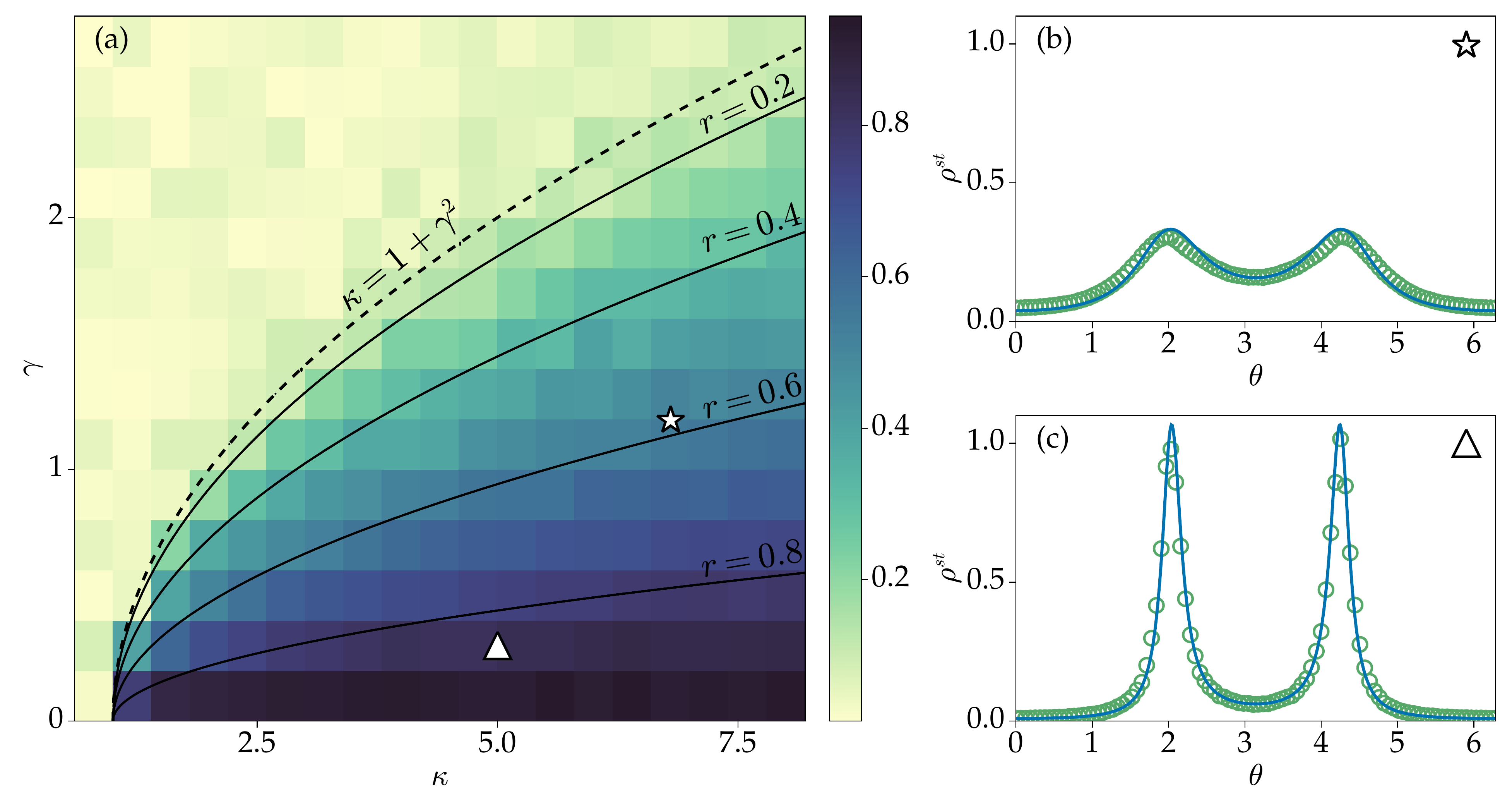}
    \caption{(a) Variation of the coherence, $r=|\lambda|$, for the stationary distribution with the stability boundary, $\kappa=1+\gamma^2$ indicated with a dashed line. Each square shows a simulation of $N=10^3$ oscillators until $t=100$ with $2\times10^4$ timesteps initialised at the incoherent state. The colour indicates the fitted value $|\lambda|$ of time-averaged stationary distribution after $t=87.5$. (b-c) Stationary distributions for Lorentz distributed frequencies and Cauchy noise calculated from the $4\times10^4$ timesteps for $N=10^4$ until $t=100$. Blue line shows the Kato-Jones distribution in \eqref{eq:stationarydistr} with parameters given by the triangle and star in (a) respectively.}
    \label{fig:lorentzstationary}
\end{figure}

Consequently, \eqref{eq:zintermsoflambda} reduces \eqref{eq:omegazlambda} to the algebraic equation
\begin{align}
    i\lorentzsig & = 1 - \frac{\kappa}{2} \frac{(\lambda+\bar{\lambda})}{1+|\lambda|^2} (\lambda+\lambda^{-1})\,.
    \label{eq:algebraiceqn}
\end{align}
Of the possible roots within the unit disc (accounting for rotation and reflection symmetry) we pick the one in the top right quadrant and write it as $\lambda=|\lambda|e^{i\Delta}$. The solution can then be explicitly stated 
\begin{align}
    |\lambda|  &= \left(\frac{\sqrt{\kappa-1}-\lorentzsig}{\sqrt{\kappa-1}+\lorentzsig}\right)^{\nicefrac{1}{2}}\,, \quad \Delta = \arccos(1/\sqrt{\kappa}).
\label{eq:coherence}
\end{align}
This solution only exists for $\kappa>\lorentzsig^2+1$, matching the stability condition for the incoherent state, and in the limit $\gamma\to0$ it recovers the Dirac mass pair solution obtained above. The argument of $\lambda$ is the separation between the peaks in $\rho$, while the modulus controls their coherence. For small values of $\gamma$ and large values of $\kappa$, the peaks are sharp whereas they become less pronounced when $\gamma$ is small and as $\kappa\to\gamma^2+1$ as can be seen in \cref{fig:lorentzstationary}. From this solution we also obtain a closed expression for the averaged coherence order parameter: $|z| = 2|\lambda| \cos(\Delta)/(|\lambda|^2+1)=\sqrt{(\kappa-\lorentzsig^2-1)/\kappa(\kappa-1)}$. 

\section{Approximate low dimensional dynamics}

We now extend the method applied above beyond the stationary states to deduce approximate low-dimensional dynamics for the evolution of the disorder averaged oscillator phase density for Cauchy noise. Applying the disorder average to \eqref{eq:rhoevolution} in the case $\alpha=1$, we obtain
\begin{equation}
\begin{aligned}
    \frac{1}{|k|}\partial_t\gav{\rho_k}  =&-(1+\gamma^2)\gav{\rho_k} + \kappa\big(  z\gav{\rho_{k-1}}+\bar{z}\gav{\rho_{k+1}} \big) \\ 
    & -\frac{\kappa^2}{4}\left(\bar{z}^2\gav{\rho_{k+2}}+2|z|^2\gav{\rho_k}+z^2\gav{\rho_{k-2}}\right)\,.
\label{eq:rhoavevolution}
\end{aligned}
\end{equation}
Similarly, the disorder average of the T\"{o}njes-Pikovsky ansatz \eqref{eq:generalansatz} is simply $\gav{\rho_k}=c\lambda^k+\overline{c\lambda^k}$. This is consistent with \eqref{eq:rhoavevolution} if $c$ is assumed constant, and if 
\begin{align}
    \dot{\lambda} = -\lambda\left[1-\kappa(c\lambda+\overline{c\lambda})(\lambda+\lambda^{-1})\right]^2-\lambda\gamma^2.
    \label{eq:approxlowdimODE}
\end{align}
This equation describes the approximate low dimensional dynamics of the order parameter $\lambda$. Unlike the Ott-Antonsen manifold for the Kuramoto model with Cauchy noise and Lorentz intrinsic frequencies, this is not an exact mapping; the coefficient $c$ actually has a non-constant imaginary part, which was ignored in the derivation of \eqref{eq:approxlowdimODE}. Nonetheless, we find it provides a good qualitative description of the evolution of the system in simulation experiments. 
\begin{figure}[t]
    \centering
    \includegraphics[width=8.2cm]{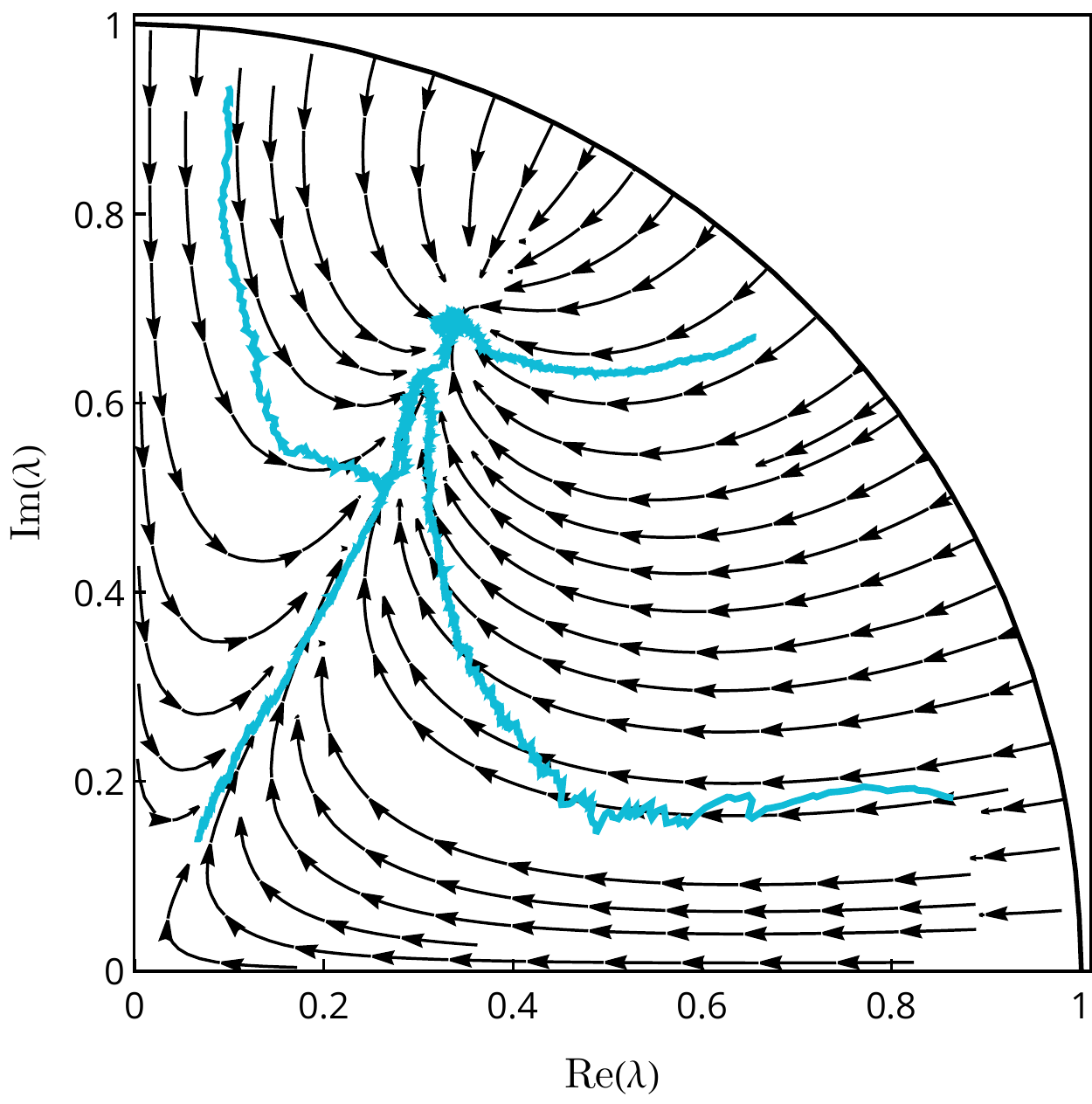}
    \caption{Field lines for the approximate low dimensional dynamics given in \eqref{eq:approxlowdimODE} with $\kappa=5,\;\gamma=0.1$ as in \cref{fig:lorentzstationary} (b). Cyan lines show paths of the fitted order parameter according to the approximate distribution from oscillator simulations.}
    \label{fig:vecfield}
\end{figure}

To test the predictive power of our approximate low dimensional dynamics, we prepare a finite sample system in a state consistent with the T\"{o}njes-Pikovsky ansatz and forward integrate. A time series for $\lambda$ can be inferred by fitting the empirical distribution of oscillator phase to the Kato-Jones distribution in \eqref{eq:stationarydistr}. Full details of how to correctly prepare the samples are provided in Appendix \ref{app:sampling}. \Cref{fig:vecfield} shows the complex flow field described by \eqref{eq:approxlowdimODE}, overlayed with simulation results for various initial values of $\lambda$.  We see that the approximate low dimensional dynamics represent well the trajectory of the order parameter as it evolves towards the steady state. 

\section{Conclusions}

To summarise, we have sought to find a model of synchronization arising from coupling purely in the noise strength on each oscillator. This model differs from almost all previous models of synchronization as the tendency towards synchrony is completely intrinsic to the system, with each oscillator acting under independent noise. In contrast, the existing literature focuses on deterministic coupling or random coupling through external or common noise to each oscillator. 
The specific choice made reproduces the exact stability condition about the incoherent state for the noisy Kuramoto model with a general frequency distribution \cite{strogatz1991}. 
For the other stationary state, comparisons can be made to systems with multi-harmonic, deterministic coupling \cite{cauchynoise}. 
We applied the approaches developed for such systems to this model, which enabled us to characterise the binary synchronised steady state in terms of a Kato-Jones distribution. 
While not being exact, this description was also useful in describing the general dynamics of the system in terms of the order parameter $\lambda$. 
It remains to be seen if an exact description of the low dimensional manifold can be found. 
This has also been a challenge for more traditional Kuramoto models with noise. 
Recent developments have shown that the Ott-Antonsen ansatz can be generalised to a larger family of invariant manifolds \cite{beyondOtt} and so it is possible a similar approach could be taken for the system present in this article.

{\it Acknowledgements.} JW supported by the EPSRC: EP/S022945/1. 

%

\appendix

\section{Stability of the incoherent state}
\label{app:stabincoherent}

In terms of the density of oscillators, $\rho(\theta,\omega,t)$, the general system in the main text is
\begin{align}
    \partial_t\rho = - \omega\partial_\theta \rho+\partial^{\alpha}_{|\theta|}\left(\rho\,\gav{\big.\rho \ast f }^2 \right).
    \label{eq:generaldensityevolutionApp}
\end{align}
We study the linear stability about the incoherent state by writing $\density(\theta,\omega,t) = \rho_\circ + \varepsilon\fluc(\theta,\omega,t)$ for small $\varepsilon>0$ and where $\rho_\circ=1/2\pi$. Substituting this into \eqref{eq:generaldensityevolutionApp}, we have that 
\begin{subequations}
\begin{align*}
    \varepsilon\partial_t\fluc & = \partial_{|\theta|}^\alpha \left[\left(\rho_\circ + \varepsilon\fluc\right)\gav{\left(\rho_\circ+\varepsilon\fluc\right)\ast f}^2\right] -\omega\varepsilon\partial_\theta\fluc \\
    & = \partial_{|\theta|}^\alpha \left[(\rho_\circ+\varepsilon\fluc)\left(f_0 + \varepsilon\gav{f\ast\fluc}\right)^2\right] -\omega\varepsilon\partial_\theta\fluc \\
    & = \partial_{|\theta|}^\alpha \left[2\varepsilon\rho_\circ f_0\gav{f\ast\fluc} + \varepsilon f_0\fluc\right] -\omega\varepsilon\partial_\theta\fluc + \mathcal{O}(\varepsilon^2) 
\end{align*}
\end{subequations}
where $f_0=\frac{1}{2\pi}\int_{-\pi}^\pi f(\theta)\d{\theta}$. Therefore, the linearised fluctuations about the incoherent state evolve according to
\begin{subequations}
\begin{align}
    \partial_t\fluc & =  f_0\partial_{|\theta|}^\alpha\left(\fluc +2\rho_\circ\gav{f\ast\fluc}\right) -\omega\partial_\theta\fluc \\
    & = f_0\partial_{|\theta|}^\alpha\fluc +2\rho_\circ f_0\gav{(\partial_{|\theta|}^\alpha f)\ast\fluc}-\omega\partial_\theta\fluc.
\end{align}
\end{subequations}
Similar stability conditions have been found for systems with deterministic coupling and independent brownian noise. Strogatz and Mirollo \cite{strogatz1991} studied the Kuramoto model with stochastic noise and general intrinsic frequencies. Here we follow their approach with the generalised Levy noise used above. The evolution of the oscillators with Kuramoto coupling and Levy noise is 
\begin{align}
    \dot{\theta}_n= \omega_n + \frac{K}{N}\sum_{m=1}^n\sin(\theta_m-\theta_n) + \xi_n(t).
\end{align}
For this system, we can write the density of oscillators as 
\begin{align}
    \partial_t \rho =  \partial_{|\theta|}^\alpha\rho  - \omega\partial_\theta\rho + K\partial_\theta(\rho\gav{\sin \ast \rho})
\end{align}
Again, applying the linear perturbation about the incoherent state, we obtain
\begin{subequations}
\begin{align*}
    \partial_t \fluc & =  \partial_{|\theta|}^\alpha\fluc + K\partial_\theta\left(\rho_\circ\gav{\sin \ast \fluc}+\fluc\gav{\sin\ast\rho_\circ}\right) - \omega\partial_\theta\fluc \\
    & = \partial_{|\theta|}^\alpha\fluc + K\rho_\circ\partial_\theta\gav{\sin \ast \fluc} - \omega\partial_\theta\fluc \\
    & = \partial_{|\theta|}^\alpha\fluc + K\rho_\circ\gav{\cos\ast\fluc} - \omega\partial_\theta\fluc
\end{align*}
\end{subequations}
where in the second line we have used that $\int_{-\pi}^\pi\rho_\circ\sin(\theta)\d \theta=0$. Comparing this with the stability for our model, it can be seen that the stability conditions match if the coupling function is chosen such that $f_0=1$ and 
\begin{align*}
    2\partial_{|\theta|}^\alpha f(\theta) = K\cos(\theta).
\end{align*}
The form of the function $f$ is apparent if we consider its Fourier modes:
\begin{align*}
    -|k|^\alpha f_k = \frac{K}{4}(\delta_{k,-1}+\delta_{k,1}) - \delta_{k,0}
\end{align*}
and so $f_{\pm1}=-K/4$, $f_k=0$ for $k\neq\pm1,0$ and $f_0=1$ as before. Thus, the only functional form which matches for all Fourier modes is
\begin{align}
    f(\theta) = 1- \kappa\cos(\theta)
    \label{eq:cosinecoupling}
\end{align}
with $\kappa=K/2$.

\section{Sampling from the approximate low dimensional distribution}
\label{app:sampling}

To find an approximate manifold for the dynamics of the system, we proposed that the order parameters took the form 
\begin{align*}
    \rho_k(\omega) = c_1(\omega)\lambda_1(\omega)^k + c_2(\omega)\lambda_2(\omega)^k\;.
\end{align*}
To sample an initial condition, we first note that no assumption was made of the form of $\lambda(\omega)$ besides requiring analyticity of $\rho_k$ and thus $z$. In other words, we must have that $|z(\omega)|<1$. 

One possible initial condition is $\lambda_1(\omega)=Ke^{-i|\omega|\phi}$ where, for simplicity, we choose $\lambda_1=\overline{\lambda_2}$. The overall order parameter, $\gav{\lambda}$, can be found from 
\begin{subequations}
\begin{align*}
    \gav{\lambda} & = \int_{-\infty}^\infty g(\omega)\lambda_1(\omega)\d \omega \\
    & = \lambda_1(-i\gamma^2) = Ke^{-i\gamma^2\phi}\;.
\end{align*}
\end{subequations}
Thus, recalling that $\gav{\lambda}=\lambda=|\lambda|e^{i\Delta}$, we have that 
\begin{align*}
    |\lambda|=|K|e^{\im(\phi)\gamma^2},\quad \Delta=\arg(K)-\gamma^2\re(\phi)\;.
\end{align*}
Given a distribution width, $\gamma^2$, we can choose $(K,\phi)$ to give the $(|\lambda|,\Delta)$ we desire for the initial condition. Since we require $|\lambda_1(\omega)|<1$ for all $\omega$ and $|\bar{\lambda}|<1$, we are constrained to $\im(\phi)<0$. Further to this we must have $|K|<1$ which means
\begin{subequations}
\begin{align*}
    1 & > |\lambda|e^{-\im(\phi)\gamma^2} \\
    \ln(|\lambda|^{-1}) & > -\im(\phi)\gamma^2.
\end{align*}
\end{subequations}
In summary,
\begin{align*}
    \frac{1}{\gamma^2}\ln|\lambda| < \im(\phi) < 0, \quad |K|<1\;.
\end{align*}
Beyond this, the parameters are free to be chosen in any way to obtain the desired order parameter. For simplicity in our simulations we choose $\im(K)=\re(\phi)=0$.

\section{Stochastic Asymptotic Stability}
\label{app:stochstability}

By considering the noise strength on an individual particle due to the mean field of all particles, we can understand the dynamics of the system in terms of two states. The SDE of an individual particle when there are no intrinsic frequencies (and centering the distribution on zero so that $\arg(z)=0$) is 
\begin{align*}
    \dot{\theta} = \big(1-\kappa |z|\cos(\theta)\big)\xi(t).
\end{align*}
The coupling strength indicates the state each particle gravitates towards. Particles get trapped in regions with small noise strength and diffuse away faster from regions with large noise strength. The result is that, eventually, particles will tend towards the minimum of the noise strength: $\left(1-\kappa |z|\cos(\theta)\right)^2$. 

When $\kappa R<1$, the system behaves similarly to the Kuramoto model as the particles tend towards the mean phase, increasing the overall coherence. The difference comes once the coherence reaches the point that $\kappa |z|$ and two minima exist at $\pm\arccos(1/\kappa |z|)$. The particles are equally attracted to these points and eventually all particles are equally distributed between these two phases. At this state $|z|=1/\sqrt{\kappa}$ and all particles are at $\pm\Delta=\pm\arccos(1/\sqrt{\kappa})$. Initially the particles diffuse but the ones around the mean phase do so less strongly. Particles coalesce onto this region until the kernel changes and then particles either side of the mean become static. These static regions move away from the mean phase slowly as more particles condense onto the two points. Here we discuss the stability of this binary synchronised state from the perspective of the stochastic stability of a single oscillator. First, we define what it means for an oscillator to be stochastically stable.


\begin{theorem}[Stochastic asymptotic stability \cite{MaoStochStability}]
    Assume a SDE has a trivial solution $x=0$. The trivial solution is stochastically asymptotically stable (SAS) if it is stochastically stable and for every, $\varepsilon\in(0,1),\exists\delta_0 = \delta_0(\varepsilon)>0$ such that
    \begin{align*}
        \mathbb{P}\left\{\lim_{t\to\infty}|x(t;x_0)|=0\right\} \geqslant 1-\varepsilon
    \end{align*}
    whenever $|x_0|<\delta_0$.
    \label{thm:aysmstability}
\end{theorem}

Suppose the system is in the binary synchronised state with $\beta=1$ (Brownian noise). The SDE for a single stray oscillator away from the two peaks is then
\begin{align}
    \dot{\theta} = \big(1-\sqrt{\kappa}\cos(\theta)\big)\xi(t)\;.
    \label{eq:strayparticleSDE}
\end{align}
The mean first passage time, $\tau(\theta)$, for the oscillator starting in the region $[-\Delta+\epsilon,\Delta+\epsilon]$ to reach a distance $\epsilon$ from the peaks is 
\begin{align*}
    \deriv[2]{\tau}{\theta} = \left[1-\sqrt{\kappa}\cos(\theta)\right]^{-2}
\end{align*}
with the boundary conditions $\tau(\epsilon-\Delta)=\tau(\Delta-\epsilon)=0$. We can also determine which peak it is likely to join given a starting point $\theta_0$.
For an SDE with no drift, the probability, $p_i$, of exit through a boundary, $b_i$, given an initial position, $x_0$, is \cite{gardiner}
\begin{align*}
    p_1(x_0) = \frac{b_2-x_0}{b_2 - b_1}, \quad p_2(x_0) = \frac{x_0-b_1}{b_2 - b_1}\;.
\end{align*}
Thus in this case,
\begin{align*}
    p_{\pm\Delta}(\theta_0) = \frac{\Delta\pm\theta_0}{2\Delta}\;.
\end{align*}
Suppose that the particle starts a distance $\vartheta_0$ from the peak at $-\Delta$. Writing $\theta=\vartheta-\Delta$ we have that
\begin{subequations}
\begin{align*}
    \mathbb{P}\left[\lim_{t\to\infty}\vartheta(t)=0\;|\vartheta(0)=\vartheta_0\right] & = p_{-\Delta}(\vartheta_0-\Delta) \\
    & = 1 - \frac{\vartheta_0}{2\Delta}.
\end{align*} 
\end{subequations}
Therefore, from \Cref{thm:aysmstability}, the particle is stochastically asymptotically stable with $\delta_0(\varepsilon)=2\Delta\varepsilon$ and $\varepsilon = \vartheta_0/2\Delta$. If all oscillators are perturbed such that we still have $\sqrt{\kappa'}=\kappa|z|>1$ and $\arg(z)=0$, the SDE for each particle is of a similar form as \eqref{eq:strayparticleSDE}: 
\begin{align*}
    \dot{\theta}_n = \big(1-\sqrt{\kappa'}\cos(\theta_n)\big)\xi_n(t)\;.
\end{align*}
We conclude that the distribution is also SAS in the thermodynamic limit $N\to\infty$ since all perturbed particles at least appear to be SAS near the binary synchronised state.

\end{document}